\documentclass[12pt]{iopart}
\usepackage{braket}
\usepackage{graphicx}
\def\tmax{\ensuremath{{t_\mathrm{max}}}}
\def\tmin{\ensuremath{{t_\mathrm{min}}}}
\begin{document}

\title{Hawkes process as a model of social interactions: a view on video dynamics}

\author{Lawrence Mitchell$^{1,2}$ and Michael~E.~Cates$^1$}
\address{$^1$SUPA, School of Physics and Astronomy,
  and $^2$EPCC, University of Edinburgh, JCMB Kings Buildings, Mayfield
  Road, Edinburgh EH9 3JZ, United Kingdom}
\ead{lawrence.mitchell@ed.ac.uk}

\begin{abstract}
  We study by computer simulation the ``Hawkes process'' that was
  proposed in a recent paper by Crane and Sornette
  (Proc. Nat. Acad. Sci. USA {\bf 105}, 15649 (2008)) as a plausible
  model for the dynamics of YouTube video viewing numbers.  We test
  the claims made there that robust identification is possible for
  classes of dynamic response following activity bursts.  Our
  simulated timeseries for the Hawkes process indeed fall into the
  different categories predicted by Crane and Sornette. However the
  Hawkes process gives a much narrower spread of decay exponents than
  the YouTube data, suggesting limits to the universality of the
  Hawkes-based analysis.
\end{abstract}


\section{Introduction}
Recently \cite{Crane:2008}, Crane and Sornette analysed the viewing of
YouTube videos as an example of a nonlinear social system. They
identified peaks in the timeseries of viewing figures for around half
a million videos and studied the subsequent decay of the peak to a
background viewing level.  A self-excited Poisson process, or Hawkes
process \cite{Hawkes:1971}, was proposed as a model of the
video-watching dynamics, and a plausible link made to the social
interactions that create strong correlations between the viewing
actions of different people. Individual viewing is not random but
influenced by various channels of communication about what to watch
next.
  
The Hawkes process has a Poisson distributed number of
views, with an instantaneous rate given by
\begin{equation}
  \label{eq:1}
  \lambda(t) = \eta(t) + \sum_{\{t_i<t\}} \mu_i \phi(t - t_i).
\end{equation}
Here $\eta$ is a noisy source term (allowing viewing to occur even for
a completely dormant video, for instance) and the summation describes
how past viewing events at times $\{t_i\}$ influence the current event
rate.  The coefficient $\mu_i$ is the number of potential viewers
influenced directly by person $i$ who viewed a video at time $t_i$;
the function $\phi$ describes the waiting time distribution for those
influenced, between trigger and response. (Put differently, this is
the distribution of waiting times between finding out about a
particular video and actually viewing the video.).  If $\phi(t)$ is a
power-law memory function as used in this work, the resulting process
is also known as an ETAS model \cite{Helmstetter:2002}, and can be
used to model earthquake aftershocks.

On the basis of previous work \cite{Johansen:2000}, Crane and Sornette
chose a long-memory waiting time distribution
\begin{equation}
  \label{eq:2}
  \phi(t) \sim t^{-(1+ \theta)} \qquad \theta \in (0, 1).
\end{equation}
For fixed $\theta$ (we address variability in $\theta$ later on), the
behaviour of a timeseries generated by such a Hawkes process then
depends on the distribution of $\mu$.  There are four separate
dynamical classes, two if a viewing shock happens from an external
stimulus, two from internal dynamics \cite{Sornette:2003}.  In this
paper we address only the dynamics of externally shocked timeseries,
but for completeness all four dynamic classes will be outlined below.

In each dynamic class, there is a different prediction for the
power-law decay of the activity level following an initial shock. The
power laws involved are quite distinct for each class, and predicted
by \cite{Sornette:2003} to depend solely on $\theta$, whereas the
statistics of $\mu$ control solely which class one is in. Therefore,
if Eq.(\ref{eq:2}) were really to hold (with a unique $\theta$) one
might naively expect the distribution of power law exponents observed
in the data to collapse onto a set of discrete delta-functions, one
for each dynamic class. On reflection, however, this cannot be correct
since an individual activity burst represents a sequence of discrete
events which (unless the total number of these is enormously large) is
unlikely to be fully self-averaging for the purposes of fitting a
power law to the long-time decay. In practice for the YouTube data
\cite{Crane:2008} the distribution of fitted exponents is very broad
with, at best, bump-like features at the expected discrete exponent
values. Crane and Sornette get around this by using a quite separate
method (detailed below) to classify the dynamic class of each burst,
and then showing that the subdistributions for each class are unimodal
with modal values close to the expected one for that class. The
overlapping exponent distributions that necessitate this procedure do
however call into question the announced robustness \cite{Crane:2008}
of the dynamic classes inferred from the Hawkes model.

In the present work we perform simulations that shed light on how much
of this exponent variability can be expected to arise from the Hawkes
process itself. Any variability beyond this level in the YouTube data
is evidence that Eq.(\ref{eq:2}) does not really describe the social
dynamics of YouTube. Of course, nobody would expect this dynamics to
be captured {\em exactly} by the Hawkes process; however, behind the
concept of robust dynamic classes in \cite{Crane:2008} lies a broader
notion of universality. For instance in equilibrium critical
phenomena, a very simple model (the Ising model) captures to arbitrary
accuracy the universal features of a wide class of phase transitions
involving order parameters of the same symmetry. In the wider context
of nonequilibrium criticality, the universal status of simple models
is much less well established, and deserves detailed attention. Our
simulation results suggest that this universality may be somewhat
limited, at least if one is interested in the {\em distribution} of
fitted exponents for individual activity bursts within each dynamic
class.

In what follows we first classify all four dynamic regimes before
presenting the analysis of our results.

\section{External shock}
\label{sec:external-shock}

In this regime, the viewing rate is first dominated by the $\eta$ term
in Eq.(\ref{eq:2}). At some time, $t_0$, the video gains widespread
public attention. (It might be featured in a national newspaper, or on
some high-traffic website; or it may relate to a famous person whose
death is suddenly announced.)  This produces a spike in the viewing
figures which then decay away.  The form of the decay depends on the
distribution of $\mu$.
\begin{itemize}
\item If $\braket{\mu} = 1$, a cascade of viewing events occurs and
  the timeseries decays from the shock like $\sim t^{-(1-\theta)}$.
  This is termed a critical decay.
\item If $\braket{\mu} < 1$, only first generation viewing events are
  important (i.e., those stimulated by the external source) and the
  timeseries decays like $\sim t^{-(1+\theta)}$.  This is a
  subcritical decay.
\end{itemize}

\section{Internal shock}
\label{sec:internal-shock}

A particular series of viewing events can lead to an internally
created maximum in the timeseries (above that expected for a Poisson
noise process).  This internal shock has a different decay exponent
again from the externally induced peaks.  The two internal dynamic
classes are:
\begin{itemize}
\item A simple noise process if $\braket{\mu} < 1$; no coherent peak arises.
\item A peak grows and decays like $\sim t^{-(1-2\theta)}$; this occurs if
  $\braket{\mu} = 1$.
\end{itemize}

\section{Classification and exponent values}
\label{sec:ident-diff-dynam}

If this model is correct for the dynamics of video views, it should be
possible to identify the different dynamic classes by finding peaks
in the viewing timeseries and then fitting a power law to the
subsequent decay. These power laws should form a distribution which
arises as the merger of the various classes; if the individual
activity bursts can separately be classified, the subdistribution for
each class can be extracted.  Crane and Sornette perform such an
analysis and by fitting to the modal exponent for each class infer a
value for the exponent $\theta$ in Eq.(\ref{eq:2}) of $\theta \approx
0.4$.  We therefore create artificial timeseries with $\braket{\theta}
= 0.4$ for best comparison with their data.  With this choice, we
expect to extract decay exponents (recalling that we only study the
externally shocked case) of
\begin{itemize}
\item $\beta_{sc} = 1 + \theta = 1.4$ and
\item $\beta_{c} = 1 - \theta = 0.6$
\end{itemize}

As mentioned previously, the model might lead one to expect
$\delta$-function peaks in a PDF of decay exponents, corresponding to
the various dynamic classes.  The data presented in \cite{Crane:2008}
show weak peaks at these values, but with a significant spread.  In
particular, some of extracted exponents would imply values of $\theta$
that lie outside the range $0<\theta<1$ required by the model itself.
With the help of our simulation data, we can look at whether the
spread arises through miscategorisation; a poor fitting method;
fluctuations in the fitted exponents due to noise inherent in the
Hawkes process itself; or failure of the Hawkes model to accurately
describe the YouTube data.

\section{Generating the synthetic data}
\label{sec:gener-synth-data}

We carry out a discrete-time simulation of the Poisson/Hawkes process,
restricting attention to activity bursts initiated by external
shocks. (We generally take each initial shock to comprise $N_0 = 5000$
views.)  We choose to generate a random number of views from a Poisson
distribution with given mean at each timestep and update the rate
accordingly afterward.  Effectively, we treat the continuously varying
$\lambda(t)$ as a constant, generate a given number of events and then
modify $\lambda$ according to Eq.~(\ref{eq:1}).  We must also choose
what values $\theta$ and $\mu$ can take.  Additionally, we need a
normalisation for the distribution of waiting times, $\phi$.
Following \cite{Crane:2008} we take this distribution normalized to
unity (so that all those `influenced' to watch a video by a particular
viewing event do watch it eventually). Remembering that the waiting
time will be an integer (due to our simulation strategy), we have
\begin{equation}
  \label{eq:3}
  \phi(t) = \frac{1}{t^{1 + \theta} \zeta(1+\theta)}
\end{equation}
with $\zeta$ the Riemann $\zeta$-function. This ensures
\begin{equation}
  \label{eq:4}
  \sum_{t=1}^\infty \phi(t) = 1.
\end{equation}

Our algorithm for generating the synthetic data is therefore as
follows:
\begin{enumerate}
\item Shock the system by creating $N(0) = N_0$ initial viewing
  events.
\item For each viewing event, generate the number of subsequent
  viewers $\mu_i$ by sampling from $P(\mu)$.  At time $t$ we therefore
  seed $n = \sum_{i=1}^{N(t)} \mu_i$ future viewing events.  Each of
  these $n$ future viewers has their own decay constant $\theta_i$
  drawn from $P(\theta)$.
\item Generate a Poisson event rate $\lambda(t)$ by summing over the
  past history according to Eq.~\ref{eq:1}.
\item Use this rate to generate the number of viewing events $N(t)$
  between time $t$ and $t + \delta t$.
\item Increment $t$ by $\delta t$ and repeat steps (ii) to (v) until
  the maximum specified time has been reached.
\end{enumerate}

The model analysis in \cite{Crane:2008} assumes that all $\theta$
values are equal.  It is, however, not clear that all interactions
would involve exactly the same response kernel.  If those influenced
have a distribution of waiting habits, this can safely be averaged
unless there is a correlation with the person exerting the influence
(so that $\theta$ varies between the events $i$ in Eq.(\ref{eq:2})). To
allow for the latter possibility we carry out simulations not only
with a single $\theta = 0.4$ but with a random distribution of
$\theta_i$ to see if this modifies the results.  For the latter we
choose $\theta$ from a truncated Gaussian with mean $0.4$ and standard
deviation $0.2$ (restricting the support to $\theta \in (0,1)$).

Finally we need to choose the statistics of $\mu_i$.  We will see that
the particular choice of distribution does not make an appreciable
difference to the results for the externally shocked system (although
as detailed above, the value of $\langle\mu\rangle$ is important).
Here we present results where $\mu$ is drawn from appropriately
weighted $\delta$-function distributions as well as Poisson
distributions.

\section{Fitting the data}
\label{sec:fitting-data}

We estimate decay exponents from the artificially produced timeseries
both via the method described in \cite{Crane:2008} and using a maximum
likelihood estimator.  The least squares estimator used in
\cite{Crane:2008} can give incorrect parameters \cite{Clauset:2007}
since some of the assumptions behind it are violated for power law
decays.  However, in our study we find little difference between the
maximum likelihood decay exponents and the least squares exponents,
which is evidence that errors in the exponent estimation method used
in \cite{Crane:2008} are not the main cause of the large spread of
observed exponents.

\subsection{Maximum likelihood estimator}
Each post-shock timeseries decay has two free parameters once the peak
has been identified: the decay exponent $\beta$, and the time at which
the peak has decayed to the background noise level $\tmax$.  To
construct the maximum likelihood estimator (MLE) for $\beta$, we
assume the data to be independent identically distributed random
variables drawn from a discrete power law distribution, $P(t, \beta)$.
That is, with a peak occurring at $t=0$, we expect, for $t = 1, \dots,
\tmax$
\begin{equation}
  \label{eq:6}
  P(t, \beta) = \frac{t^{-\beta}}{H_{\tmax, \beta}}
\end{equation}
where 
\begin{equation}
  \label{eq:7}
  H_{\tmax, \beta} = \sum_{k=1}^\tmax k^{-\beta}
\end{equation}
is a generalized harmonic number.

For every possible \tmax\ we find the best fit value of $\beta$ for
this distribution using a maximum likelihood estimator.  For a
timeseries $A(t)$, our dataset consists of $A(t_i)$ observations at
each time $t_i$.  Each of these $t_i$'s has an individual likelihood
given by $P(t_i, \beta)$.  We assume each observation of $t_i$ is independent
and so the likelihood of the dataset factorises into the product of
the individual likelihoods
\begin{equation}
  \label{eq:8}
  \mathcal{L(\beta)} = \prod_{\{t_i \leq \tmax\}} P(t_i, \beta).
\end{equation}
To find the best fit value of $\beta$ for a given dataset, we maximise
the likelihood with respect to $\beta$.  (In fact, since the
likelihood is such a small value, we instead maximise the logarithm of
the likelihood, but this gives the same result.)  To find the best
value of \tmax\ we follow the method of \cite{Clauset:2007} and choose
that \tmax\ which minimises the Kolmogorov-Smirnov distance statistic.
That is, for each value of \tmax\ we find the best fit decay exponent
and calculate
\begin{equation}
  \label{eq:5}
  D = \max_{x} | E(x) - C(x) |
\end{equation}
where $E(x)$ is the empirical cumulative distribution function, and
$C(x)$ the best-fit-hypothesis cumulative distribution function.  We
then pick \tmax\ as that value which minimises $D$.  (Note that this
fitting method, while finding the best fit, tells us nothing about the
quality of that fit.)

\subsection{Least squares
  estimator}
\label{sec:comp-citetcr-fitt}

We also calculate the decay exponents for the same timeseries using
the method described by Crane and Sornette \cite{Crane:2008}.  This
uses a least squares regression on the dataset to find the decay
exponent from the peak over a fixed window size.  For each fit, they
look at the distribution of the relative residuals, i.e., the
difference between the model and the empirical data, divided by the
expected value at that point.  If the relative residuals are not
distributed normally, the fit for that window size is rejected.  The
best fit to the data is chosen to be the largest window size with
normally distributed relative residuals.  Following \cite{Crane:2008}
we reject the fit if the hypothesis of a normal distribution is
violated at the 1\% level using a $\chi^2$ test.

\subsection{Fitting to an ensemble average}
\label{sec:fitting-an-ensemble}

The individual timeseries that we generate are subject to a reasonable
amount of noise giving a spread of best fit decay exponents.  Given
that we control the time and size of the initial shock, we can easily
obtain better statistics for the different parameter choices by
considering ensemble averaged timeseries.  This allows us to observe
the behaviour of the decay for a longer time and get a better fit for
the decay exponents.

The fitting method in this case is as before; we obtain the best fit
$\beta$ value by maximising the likelihood of the data.  The decay now
occurs over the whole tail of the timeseries and so we do not need to
find \tmax; we can set it manually as equal to the largest time in our
dataset.  We fit both from the peak of the decay and the `tail'.  To
determine where the tail of the data starts, we follow the same
procedure as detailed above for finding \tmax, only this time we apply
it to find \tmin.  That is, for each \tmin\ value, we calculate $D$
(Eq.~\ref{eq:5}) of the best fit and subsequently choose as our lower
cut-off that \tmin\ which minimises $D$.  We obtain errors on our
estimates of $\beta$ by noting that our MLE is asymptotically optimal
\cite{Cox:1974}, for $N$ observations, the variance in the estimated
value is therefore given by the inverse of the observed Fisher
information \cite{Tamhane:2000}
  \begin{equation}
    \label{eq:9}
    \mathrm{Var}(\beta) = \frac{1}{N \mathcal{J}(\beta)}
  \end{equation}
with
\begin{equation}
  \label{eq:10}
  \mathcal{J}(\beta) = -\frac{1}{N}\frac{\partial^2
    \log{\mathcal{L}(\beta)}}{\partial \beta^2}
\end{equation}
which can easily be obtained numerically.  The MLE is asymptotically
Gaussian with mean $\beta$ and variance given by Eq.\ref{eq:9} and so
confidence intervals are just the standard Gaussian ones.

\section{Results}
\label{sec:results}

\begin{figure}[htbp]
  \centering
  \includegraphics[width=12cm]{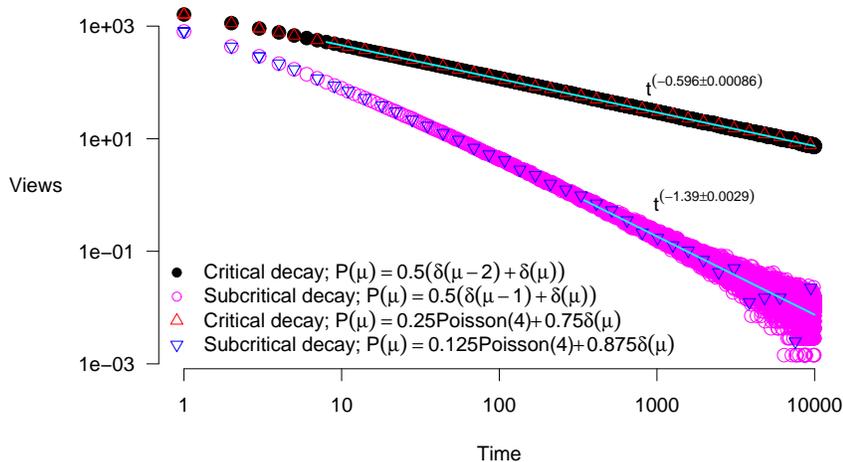}
  \caption{Ensemble average decay exponents with
    $P(\theta)=\delta(\theta - 0.4)$ and $P(\mu)$ as indicated.  Each
    dataset is the average of 700 realisations with an initial shock
    of 5000 views.  Lines show the best fit decay exponent in the tail
    of the decay ($t\geq 8$ for critical and $t\geq 344$ for
    subcritical decay) obtained from the MLE, $\pm$ figures are 95\%
    confidence intervals whose calculation is detailed in the text.
    The decay exponents are in good agreement with the theoretical
    values of 0.6 and 1.4.}
  \label{fig:mu-dist-av-ts}
\end{figure}

\begin{figure}[htbp]
  \centering
  \includegraphics[width=12cm]{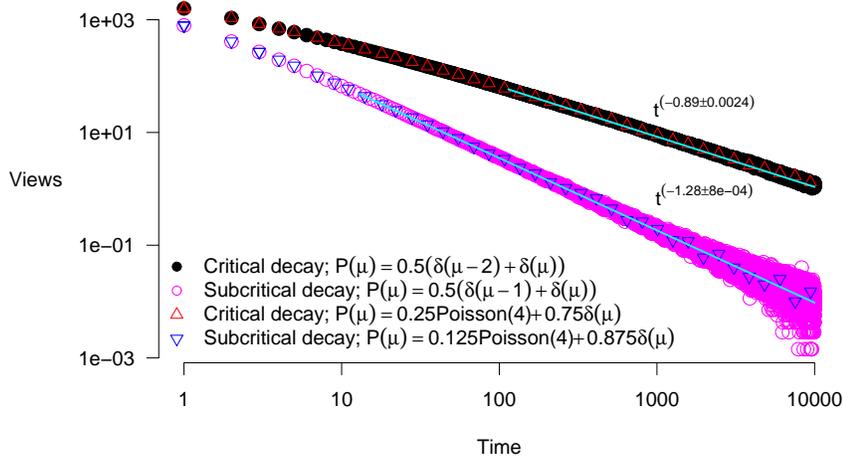}
  \caption{Ensemble average decay exponents with $P(\theta)=N(0.4,
    0.2)$ and $P(\mu)$ as indicated.  Each dataset is the average of
    700 realisations with an initial shock of 5000 views.  Lines show
    the best fit decay exponent in the tail of the decay ($t\geq 114$
    for critical and $t\geq 14$ for subcritical decay) obtained from
    the MLE, $\pm$ figures are 95\% confidence intervals.  The decay
    exponents are no longer in good agreement with the theoretical
    values of 0.6 and 1.4.}
  \label{fig:mu-theta-dist-av-ts}
\end{figure}

To begin, we look at the behaviour of the ensembled-averaged
timeseries.  As expected, we see a clear distinction between the
subcritical decay (where $\braket{\mu} < 1$) and the critical decay
(with $\braket{\mu} = 1$).  The best fit decay exponents are also
those expected (figure \ref{fig:mu-dist-av-ts}).
The difference between critical and subcritical decays remains when we
draw $\theta$ from a Gaussian distribution (figure
\ref{fig:mu-theta-dist-av-ts}).  We do, however, notice a significant
difference from the single-valued $\theta$ case: the numerical values
of the decay exponents no longer agree well with the predicted values.

Notice how the subcritical decay appears to exhibit a crossover
between a short time ``critical'' decay exponent and long time
subcritical decay.  Increasing $\braket{\mu}$ towards the critical
value of unity moves the crossover to later and later times.
Interestingly, when $\theta$ is drawn from a Gaussian distribution,
both the critical and subcritical decays exhibit some sort of
crossover behaviour, not seen in the critical decay for single-valued
$\theta$.  This crossover can contribute to the spread of exponents in
the subcritical case, since the fitting mechanism may pick up the
early time decay.  The crossover observed is discussed in detail in
the context of the ETAS model in \cite{Helmstetter:2002,Saichev:2005}.

\begin{figure}[htbp]
  \centering
  \includegraphics[width=12cm]{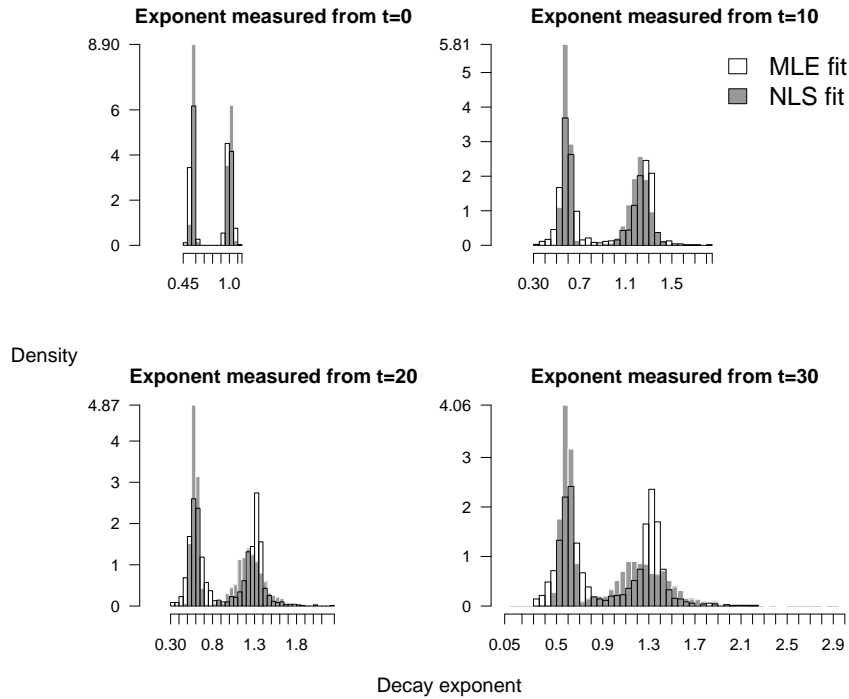}
  \caption{Histograms of extracted decay exponents for critical and
    subcritical timeseries and $P(\theta) = \delta(\theta - 0.4)$,
    initial shock is 5000, 700 realisations.  Grey histograms show
    exponents obtained from non-linear least squares fitting, white
    histograms are obtained from MLE fits.  Note how there are two
    distinct peaks in the distribution, corresponding to critical and
    subcritical decays.  The subcritical peak moves from $\beta
    \approx 1$ to $\beta \approx 1.4$ when we avoid picking up the
    early time critical decay described in the text.}
  \label{fig:mu-dist-histograms}
\end{figure}

\begin{figure}[htbp]
  \centering
  \includegraphics[width=12cm]{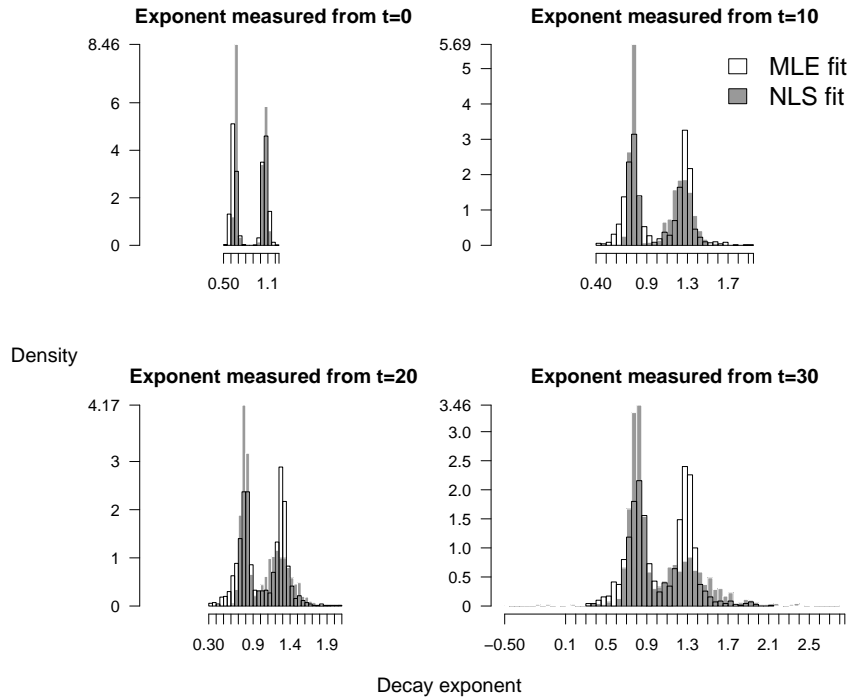}
  \caption{Histograms of extracted decay exponents for critical and
    subcritical timeseries and $P(\theta) = N(0.4, 0.2)$, initial
    shock is 5000, 700 realisations.  Grey histograms show exponents
    obtained from non-linear least squares fitting, white histograms
    are obtained from MLE fits.  Note how there are two distinct peaks
    in the distribution, corresponding to critical and subcritical
    decays.  The subcritical peak moves from $\beta \approx 1$ to
    $\beta \approx 1.3$ when we avoid picking up the early time
    critical decay described in the text.}
  \label{fig:mu-theta-dist-histograms}
\end{figure}

We now look at the distribution of decay exponents of individual
timeseries obtained from both the MLE and least squares estimator.
Our results for the ensembled averaged timeseries indicate that we
will likely not see the asymptotic exponent in the subcritical case if
we fit the entire post-shock timeseries ($\tmin = 0$), as the decay
exponent will be skewed by the early time `critical' decay.  We
therefore show results with $\tmin = 0, 10, 20$ and $30$; these latter
fits will give us an indication of what the tail exponent looks like.
The results for a single value of $\theta$ are shown in figure
\ref{fig:mu-dist-histograms}, those with $\theta$ from a Gaussian
distribution are shown in figure \ref{fig:mu-theta-dist-histograms}.

As expected, fitting the entire post-peak timeseries underestimates
the subcritical decay exponent.  Both fitting methods pick up the
early time decay, which is slower; once the early time peak is
ignored, the decay exponents are more similar to the tail seen in the
ensemble-averaged case.  We note that the fits for $\tmin \gg 1$ do
have quite a large spread of exponents.  This is due to poor
statistics in the tail of the decay: the fluctuations are large enough
that we occasionally pick up a highly anomalous decay exponent. This
form of statistical noise appears to be intrinsic to the Hawkes
process once the data is filtered by $\tmin$. Improved fits, and
presumably also narrower distributions of the fitted exponents, would
arise if we used much larger initial shocks ($N_0 \gg 5000$).

Our choice of $N_0$ is however consistent with the statements in
\cite{Crane:2008} of mean total views in the tens of thousands (with
at least 20\% of these viewed on the peak day) for the shocked
case. It is intended to give a realistic estimate of the intrinsic
noise in the Hawkes process, to see if this can account for the large
exponent spread actually found in \cite{Crane:2008}. Comparison of
their figure 4 with our figures \ref{fig:mu-dist-histograms} and
\ref{fig:mu-theta-dist-histograms} shows such an explanation to be
implausible: the exponent spread in the YouTube data is much too
large, particularly for the subcritical case.  We have also performed
simulations with $N_0 = 500$ and $N_0 = 50000$, i.e., one order of
magnitude in either direction from the results reported here.  In
studying the ensemble average of 700 such timeseries from these
simulations, we find that we cannot reject, at the 95\% significance
level, the hypothesis that the data are the same as those we have
reported for $N_0 = 5000$.  In other words, the size of the initial
shock does not affect the statistics of the resulting timeseries.  For
the small initial shocks ($N_0 = 500$), the spread of individually
fitted exponents is indeed larger than those we show here with $N_0 =
5000$ and vice versa for the larger shocks ($N_0 = 50000$).  This is
simply due to the fitting method being more (less) affected by
statistical fluctuations.  The peaks of the exponent distributions do
not, however, change appreciably.

\subsection{Classifying timeseries}
\label{sec:class-times}

Crane and Sornette do not have \emph{a priori} knowledge of which
dynamic class each timeseries belongs to.  Because the exponents do
not fall into clear classes, they use a classification method based on
the fraction of the total views that arise on the day of maximal
viewing, termed the ``peak fraction'' ($F$).  (This fraction is of
course a measure of the steepness of the subsequent decay, hence of
$\beta$.)  In our simulations, since we know $\langle \mu\rangle$, and
hence which class any given timeseries is actually in, we can look at
the peak fraction and see if this method allows for any
misclassification.  The classification according to $F$ in
\cite{Crane:2008} is to consider $F \geq 0.8$ as an exogenous
subcritical decay, $0.2 < F < 0.8$ as exogenous critical decay and $F
\leq 0.2$ as endogenous critical decay.  There are some further
comments that the classification between the two exogenous cases is
not significantly altered when varying the boundary between $F = 0.7$
and $F \rightarrow 1$.  We have not calibrated our simulations to any
of Crane and Sornette's data, and hence do not know how long the time
increment in our updates is relative to their data.  The boundaries we
choose for classification will therefore not have the same numerical
values; this will not, however, invalidate our study of the
classification method.  We find that our simulated timeseries show two
well-defined peaks in the distribution of peak fractions.  Choosing a
cut-off of $F<0.2$ to define exogenous critical and $F>0.2$ to define
exogenous subcritcal decay (recall we do not treat the endogenous
case) results in no misclassification.

\begin{figure}[htbp]
  \centering
  \includegraphics[width=12cm]{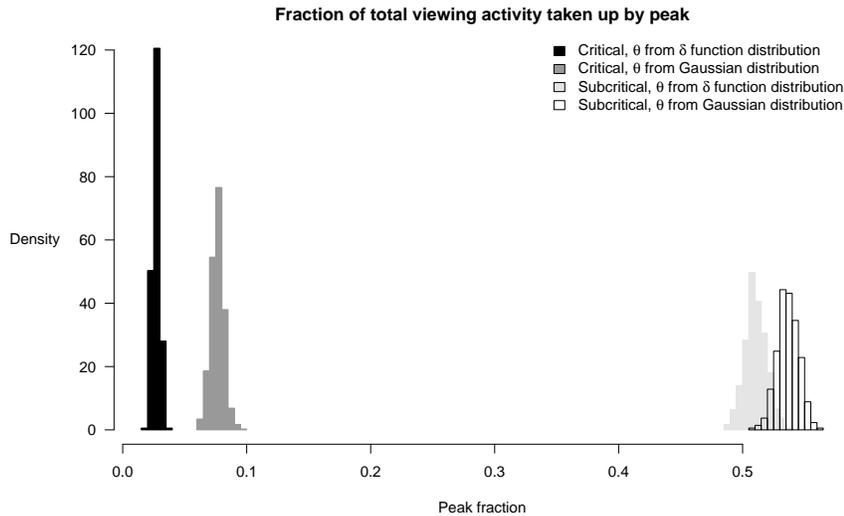}
  \caption{Distribution of peak fractions for critically and
    subcritically decaying timeseries.  Parameter values as indicated
    in legend.  Choosing a cutoff value of $F=0.2$ for the peak fraction
    would result in no misclassifications.}
  \label{fig:peak-fraction-histograms}
\end{figure}
Figure \ref{fig:peak-fraction-histograms} shows a histogram of peak
fractions of simulations with $P(\theta) = \delta(\theta - 0.4)$ and
with a Gaussian distribution.  In both cases, there is an obvious
divide between subcritically decaying timeseries and critical decays.
For a suitably placed boundary between high and low peak fractions
($F=0.2$), this method correctly classifies every single timeseries we
have studied.

\section{Conclusions}
\label{sec:conclusions}
The observed behaviour of the Hawkes process subject to external shock
is, for the case of a single-valued $\theta$ distribution, exactly as
predicted in Refs.\cite{Crane:2008,Sornette:2003}.  When $\theta$ is
drawn from a broad distribution, the numerical values of the decay
exponent are modified, but the overall picture of critical and
subcritical decays remains.  Our results show a significant spread of
fitted decay exponents, though much less than that seen in the YouTube
data reported in \cite{Crane:2008}.  We can, however, shed some light
on this.  We have good control of all the timeseries we fit to, in
particular, we ensure that they are all subject to the same size of
fluctuation (by always studying the same size of shock).  Crane and
Sornette do not have this luxury.  Our fitting to the tails of
individual timeseries shows that the exponent can vary widely if the
statistics are poor (in some instances the best-fit exponent is very
different from that of the underlying distribution of which a given
timeseries represents a single sample).  It seems likely that some of
the breadth in the range of exponents seen in \cite{Crane:2008} is
caused by considering many timeseries with poor statistics in the
tail.  By only considering timeseries with particularly large peaks
(relative to the background viewing rate), a set of decay exponents
with lower variance might be obtained. Of course, this would have a
cost in terms of the overall statistics of the sample.

In addition, our study shows that the peak fraction classification
method is a good one and we suggest that carrying out this
classification and then fitting to the ensemble average of suitably
normalised timeseries may give the best estimate of the decay
exponents.  We have also shown that the least squares fitting method
gives results that are not very different from the maximum likelihood
approach favoured here.

Our results demonstrate a way to test if $\theta$ is really a unique
global constant (equivalently, drawn from a $\delta$-function
distribution).  The ensemble-averaged timeseries in this case are
measurably different from those where $\theta$ is broadly distributed
with the same mean. Particularly, we observe a crossover effect in the
critical decay for a broad $\theta$ distribution that is not present
if $\theta$ is constant.  If the timeseries can be correctly
classified using the peak-fraction method, an ensemble average of
(suitably normalised) critical timeseries might be diagnostic of
whether $\theta$ is effectively constant or not.

Finally, we reiterate that although our analysis of synthetic Hawkes
process data results in a spread of fitted exponents within each
dynamic class, this intrinsic noise does not fully account for the
much wider distributions seen in the YouTube data of Crane and
Sornette \cite{Crane:2008}. This suggests limits to the universality
concept which presumably underlies attempts to classify activity
bursts in social systems into `robust dynamic classes'
\cite{Crane:2008}. While the Hawkes process is clearly useful in
analysing real-world data from complex social systems, some fairly
basic observables, such as the variance of the exponent distribution
for individual activity bursts, are seemingly not captured by
it. These aspects are thus either nonuniversal or lie in the
universality class of a more complex model than Hawkes.

\section{Acknowledgements} We thank EPSRC EP/E030173 for funding. MEC
holds a Royal Society Research Professorship.

\end{document}